# FinBTech: Blockchain-Based Video and Voice Authentication System for Enhanced Security in Financial Transactions Utilizing FaceNet512 and Gaussian Mixture Models

*Prof N.Jeenath Laila ME,*
*Assistant Professor,*
*Department of Computer science and Engineering,*
*Government College of Engineering,*
*Tirunelveli-7*

*Dr G.Tamilpavai ME, PhD,*
*Professor(CAS),*
*Department of Computer science and Engineering,*
*Government College of Engineering,*
*Tirunelveli-7*

## Abstract

In the digital age, it is crucial to make sure that financial transactions are as secure and reliable as possible. This abstract offers a ground-breaking method that combines smart contracts, blockchain technology, FaceNet512 for improved face recognition, and Gaussian Mixture Models (GMM) for speech authentication to create a system for video and audio verification that is unmatched. Smart contracts and the immutable ledger of the blockchain are combined to offer a safe and open environment for financial transactions. FaceNet512 and GMM offer multi-factor biometric authentication simultaneously, enhancing security to new heights. By combining cutting-edge technology, this system offers a strong defense against identity theft and illegal access, establishing a new benchmark for safe financial transactions.

**Keywords :** Blockchain, Video authentication, Voice authentication, Smart contracts, Financial transactions, FaceNet512, Multi-factor authentication, Security, Trust.

## 1.Introduction

With the introduction of digital transactions, the landscape of global finance has experienced a significant upheaval. Although this change has offered extraordinary convenience, it has also created new security risks. We present a novel integration of technologies in response to these evolving threats, utilizing the strengths of blockchain, smart contracts, InterPlanetary File System (IPFS), Advanced Encryption Standard (AES) and advanced biometric authentication using FaceNet512 and GMM. This introduction gives thorough investigation of a ground-breaking speech and video authentication system created to strengthen the security of financial transactions in the digital era.Financial transactions are carried out online on an unprecedented scale in today's linked world. However, this digital transition has made financial systems more vulnerable to a wide range of risks, such as fraud, data breaches, and identity theft. A paradigm shift in how authentication and secure financial activity is required in light of the fact that traditional security methods have failed to adequately protect these transactions.

The approach is built around the blockchain, a decentralized, unchangeable ledger.

Blockchain technology provides unmatched security via distributed consensus mechanisms and cryptographic hashes. Trustless transactions are made possible by smart contracts, executable code that is stored on the blockchain that automates and secures financial deals.

Data integrity and confidentiality are essential for the security of financial transactions. To offer the highest level of protection for sensitive data, AES encryption is used. By securely storing transaction-related files in a decentralized, tamper-resistant way, IPFS augments this encryption and lowers the danger of data breaches.

An effective and scalable method for storing and retrieving transaction records within the blockchain is provided by key-value pair storage. By streamlining data management, this ground-breaking data structure boosts authentication system speed and effectiveness while preserving the blockchain's integrity.

The solution uses FaceNet512 and GMM for cutting-edge biometric authentication. FaceNet512 uses deep learning to produce extremely accurate facial recognition, while GMM focuses on voice authentication to provide multi-factor security that is both strong and easy to use. The goal is to redefine the bar for security and trust in financial transactions by utilizing these cutting-edge technologies, offering a strong barrier against fraud and data breaches in the world of digital finance.

The paper presents the following important contributions:

1. Storing user registration information, together with audio and video inputs, on IPFS, then putting the generated content ID into Blockchain storage.
2. The "Proof of Fair Chance" consensus mechanism is introduced for transaction validation and verification.
3. Using Facenet512 and GMM to implement user voice and video authentication for system access during financial transactions.
4. Improving the effectiveness of data retrieval for a user's information from the Blockchain by using key-value pair storage.

The paper is structured as follows: Section 2 offers an in-depth review of prior research in the fields of blockchain data storage, retrieval methods, and techniques for video and voice authentication. Section 3 delves deeply into the privacy-preserving storage and retrieval model, as well as video and voice authentication techniques. It provides a comprehensive explanation of the system's architecture and the data processing flow for storage, retrieval, and authentication methods. Section 4 focuses on evaluating data storage and retrieval metrics, as well as metrics related to video and voice recognition. This section thoroughly assesses the model's performance. Finally, Section 5 summarizes the research findings and outlines potential directions for future research.

## 2.Literature Review

Ch. Rupa et al designed an Industry 5.0 based blockchain application to manage medical certificates using Remix Ethereum blockchain [1].This application also employs a Distributed Application (DApp) that uses a test RPC-based Ethereum blockchain and user expert system as a knowledge agent. The main strength of this work is the maintenance of existing certificates over a blockchain with the creation of new certificates that use logistic Map encryption cipher on existing medical certificates while uploading into the blockchain. M. Ali et al. presents a decentralized alternative to traditional Internet infrastructure, [2] which uses a combination of blockchain technology, peer-to-peer networking, and cryptographic algorithms to provide security and reliability. Mohamed Rizwan et al.'s research [3] underscores the recession resilience of Petroleum Retail Outlets, emphasizing effective management practices and addressing challenges like compliance, competition, and evolving consumer preferences. Their insights inform strategies for enhancing privacy-preserving blockchain storage and retrieval in India's fuel retail sector.

V. Rathore et al. propose a secure storage and retrieval system for encrypted data in IPFS using Shamir's Secret Sharing scheme[4] . The system is designed to provide secure storage and retrieval of data in IPFS, a peer-to-peer network for storing and sharing files. The system encrypts the data using AES and then divides it into multiple shares using Shamir's Secret Sharing scheme. These shares are then distributed across the network, making it difficult for an attacker to retrieve the original data without possessing a sufficient number of shares. Addressing fuel-related environmental concerns. Mohamed Rizwan et al.'s comparison of conventional and alternative vehicles underscores the need for transitioning to alternative fuels, aligning with privacy-preserving blockchain storage and retrieval to enhance data security and transparency in the transportation sector's sustainable evolution[5].

M. Afzal et al. propose a system for secure data sharing in IPFS using blockchain-based identity management. The system uses a combination of IPFS and blockchain technology to provide secure and decentralized storage and sharing of data. The advantages of this system include its high level of security, as the combination of AES encryption, Shamir's Secret Sharing, and blockchain-based identity management provides strong protection against attacks[6].

Blockchain provides a platform for decentralization and trust in various applications such as finance, commerce, Internet of Things (IoT), reputation systems and healthcare.However, winning demanding situations like scalability, resilience, security and privateness are but to be overcome.Due to rigorous regulatory constraints which includes Health Insurance Portability and Accountability Act (HIPAA) , blockchain packages withinside the healthcare enterprise commonly require greater stringent authentication, interoperability, and record-sharing requirements. This article [7] presents an extensive study to showcase the significance of blockchain technology from both application and technical perspectives for the healthcare domain. Abdurrashid et al. [8] gives a systematic review of blockchain scalability. It follows a scientific procedure to analyze the studies on blockchain scalability and evaluate its country of the art. It reviews the various proposed solutions and methods for blockchain scalability.

Ren et al [9] proposed a method, Dual Combination Bloom Filter (DCOMB), combining the data stream of the IoT with the timestamp of the blockchain, to improve the versatility of the IoT database system. N. R. Pradhan et al. in [10] A Google Cloud Platform-based multi-organizational, multi-host, off-chain and on-chain framework for keeping track of patient medical information as well as various peer-based plans for a hyper ledger fabric-enabled medical system that addresses the issues of data privacy, data availability, and data security have been proposed. Edward Tijan et al in [11] researches decentralized data storage represented by blockchain technology and the possibility of its development in sustainable logistics and supply chain management.

Face recognition technology is widely used in various fields, such as time and attendance, payment, access control, etc., providing great convenience to life. Siyao Qui et al proposed a face recognition model based on MTCNN and Facenet, as traditional face recognition systems mostly use manual feature setting, which has disadvantages such as low recognition accuracy and slow speed. The MTCNN model consists of three layers of convolutional neural networks, namely P-Net, R-Net and O-Net, which are used to extract faces from images. The Facenet model is used for face feature vector extraction, and Triplet Loss is used as the loss function to determine the similarity of features through the comparison of Euclidean spatial distances.[12]

Giang-Truong Nguyen and et al presented a survey of many variants of consensus algorithms inside Blockchain,[14] which could be categorized into two main types. The first type is the proof-based consensus algorithm, which is often used in public Blockchain. The second type is the voting-based consensus algorithm, which is often used in private and consortium Blockchain.To overview and provide a basis of comparison for further work, a set of performance evaluation criteria is identified . These criteria are classified into four categories including algorithm's throughput, the profitability of mining, degree of decentralization and consensus algorithms vulnerabilities and security issues. Seyed Mojtaba Hosseini Bamakan et al. systematically analyzed [15] the pros and cons of consensus algorithms and compared them in order to provide a deep understanding of the existing research challenges and clarify the future study directions.

Based on the literature review, it has been observed that there are various challenges in existing storage technology. These challenges include data privacy, scalability, reliability and slow retrieval time. These issues can negatively impact the blockchain's overall performance. The proposed system overcomes these challenges to enhance performance by providing high data privacy and fast retrieval time.

## 3. Proposed System

Figure 1 explains the general working of storing a file in IPFS and file hash to local blockchain environment Ganache.

The process begins with the Dapp creator compiling a smart contract, laying the foundation for the ensuing steps. The smart contract is subsequently migrated to the Ganache blockchain, ensuring its integration into the blockchain environment. To enable interaction between the Dapp creator and the blockchain, a connection is established through web3. Similarly, a connection between Ganache and the blockchain is established using the same web3 framework. With the Dapp in action, users gain the ability to upload files. These uploaded files are then published onto IPFS, producing a unique file hash. This hash is incorporated into the smart contract, binding the uploaded file to the blockchain. To finalize this process, transaction approval is secured, paving the way for users to conveniently view or download the associated files as needed.

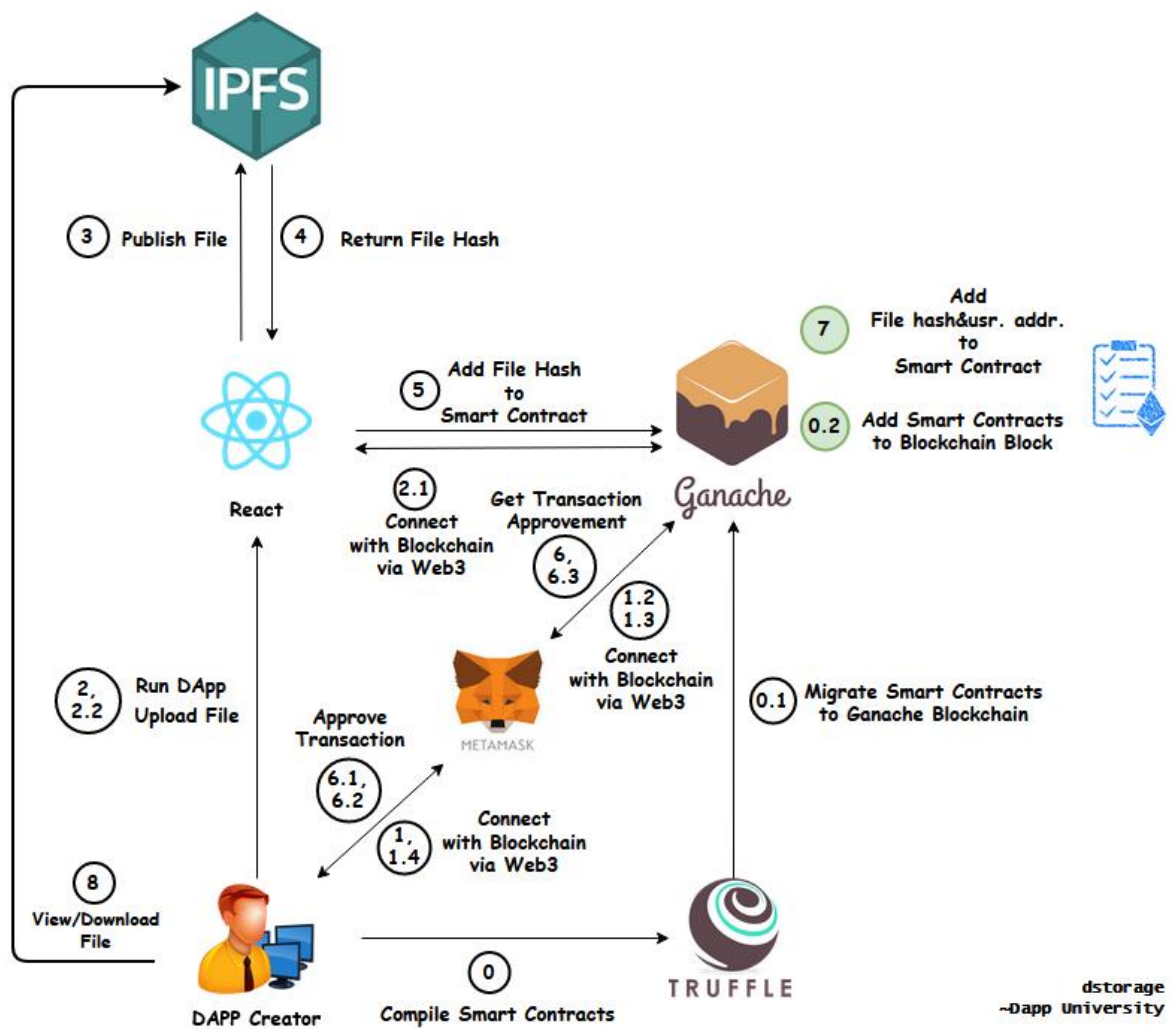

**Figure 1 : Implementation of encrypted file storing service, using blockchain and IPFS**

The proposed system, illustrated in Figure 2, is composed of the following key components:

**1. Innovative Data Storage within Blockchain:** The paper introduces a pioneering method for data storage within the blockchain, harnessing technologies such as IPFS and RPH tree. The data storage framework encompasses the following steps:

**i. User Registration in FinBTech Blockchain:** Users who intend to engage in financial transactions must first register in the FinBTech blockchain network. This registration process includes capturing audio and video inputs from the user and storing all user details in a file.

**ii. Storing Voice and Video file in IPFS:** The file is then stored in IPFS, and the Content Identifier (CID) returned by IPFS is recorded in the blockchain.

**iii. Storing Transaction in Blockchain:** Transactions are stored in the blockchain as a key value pair for efficient retrieval.

      **iv. Userid-Transaction Hash Mapping:** The smart contract is designed to link userIDs with their respective transactionHashes on the blockchain, enabling efficient transaction retrieval for users.

**2.Equitable Consensus Mechanism - Proof of Fair Chance:** The paper introduces a novel consensus mechanism designed to ensure that every miner has an equal opportunity to mine transaction data. This mechanism is aimed at preventing the centralization of mining power.

**3.User Voice and Video Authentication:** During financial transactions, user authentication occurs as follows:

      **i.User Login:** When a user logs into the FinBTech blockchain network, they are required to provide live video input and read a provided paraphrase for voice input.

      **ii.Video Authentication:** Facenet512 is employed for video authentication, comparing the live video captured during the login session with the video input provided by the user during registration. If the comparison exceeds a predefined threshold value, the user is prompted to read a paraphrase displayed on the screen.

      **iii.Voice Authentication:** After the video of the user is authenticated, the user reads the paraphrase, and their voice is recorded. Voice authentication is performed using GMM, comparing the recorded voice during the login with the voice input provided by the user during registration. Access is granted if the comparison exceeds the threshold value.

### 3.1 Innovative Data Storage within Blockchain

      The paper introduces an innovative approach to store data within the blockchain, incorporating cutting-edge technologies like IPFS for decentralized file storage and smart contracts for automating and securing various processes within the blockchain ecosystem.

### 3.1.1 User Registration in FinBTech Blockchain

      Users are often needed to supply a variety of data during the registration process in the FinBTech blockchain network, including:

**Personal Information:** This could contain the user's name, date of birth, email and phone number.

**Biometric Information:** Users are prompted to submit audio and video inputs, which are essentially recordings of their voice and facial features, as part of the biometric data collection process. For authentication purposes, these biometric data points are utilized to make sure that the individual logging in corresponds to the registered user.

---

**function userRegistration()**

```
// Prompt user to provide personal information
name = getUserInput("Enter your full name: ")
dateOfBirth = getUserInput("Enter your date of birth (YYYY-MM-DD): ")
email = getUserInput("Enter your email address: ")
phoneNumber = getUserInput("Enter your phone number: ")
// Capture audio and video inputs from the user
audioData = captureAudio()
videoData = captureVideo()
```

```
// Store user details and encrypted data on IPFS
userRecord = {
    "userID": userID,
    "name": name,
    "dateOfBirth": dateOfBirth,
    "email": email,
    "phoneNumber": phoneNumber,
    "Audio": audioData,
    "Video": videoData
}
```

### 3.1.2 Storing Voice and Video file in IPFS

The audio and video files are stored on IPFS and the CID returned by IPFS for each file is stored in the audio and video attributes within the user record. This information in the user record is then recorded as a transaction on the blockchain. This approach preserves user data privacy and minimizes the on-chain storage of actual file content.

**function storeAudioAndVideoInIPFS(audioFile, videoFile)**

```
    // Upload the audio file to IPFS and get the CID
    audioCID = IPFS.upload(audioFile)
    // Upload the video file to IPFS and get the CID
    videoCID = IPFS.upload(videoFile)
    return (audioCID, videoCID)

function UpdateUserRecordWithCIDs(userId, audioCID, videoCID):
    // Update the user record with the user's ID and the CIDs of their audio and video files
    userRecord.Audio=audioCID,
    userRecord.Video=videoCID.
    }
```

### 3.1.3 Storing Transaction in Blockchain

Smart contract is created to store a transaction containing Transaction ID, Sender Address, Transaction Timestamp, Block Number, Transaction Data containing the userRecord, Nonce, and Previous Hash with the transaction hash as the key and the remaining data as the value on a blockchain.

***Function  StoringTransactionBlockchain***

```
// Define the TransactionStorage contract
// Define the Transaction struct
    struct Transaction {
        address senderAddress;  // Address of the sender
```

```
    uint256 timestamp;      // Timestamp of the transaction
    uint256 blockNumber;    // Block number where the transaction is included
    bytes32 dataHash;       // Hash of userRecord or other data
    uint256 nonce;          // Nonce associated with the transaction
    bytes32 previousHash;   // Previous hash
}
// Declare a mapping to store transactions as key-value pairs
mapping(bytes32 => Transaction) public transactions;
// Declare a mapping of UserRecord to Userhash
mapping(bytes32 => UserRecord) public userRecords;
// Define a function to store a transaction
function storeTransaction(
    bytes32 transactionHash,
    address _senderAddress,
    uint256 _timestamp,
    uint256 _blockNumber,
    bytes32 _dataHash,
    uint256 _nonce,
    bytes32 _previousHash
)
    // Create a new Transaction struct
    Transaction memory newTransaction = Transaction({
        senderAddress: _senderAddress,
        timestamp: _timestamp,
        blockNumber: _blockNumber,
        dataHash: _dataHash,
        nonce: _nonce,
        previousHash: _previousHash
    });
    // Store the transaction with the given hash in the transactions mapping
    transactions[transactionHash] = newTransaction;
```

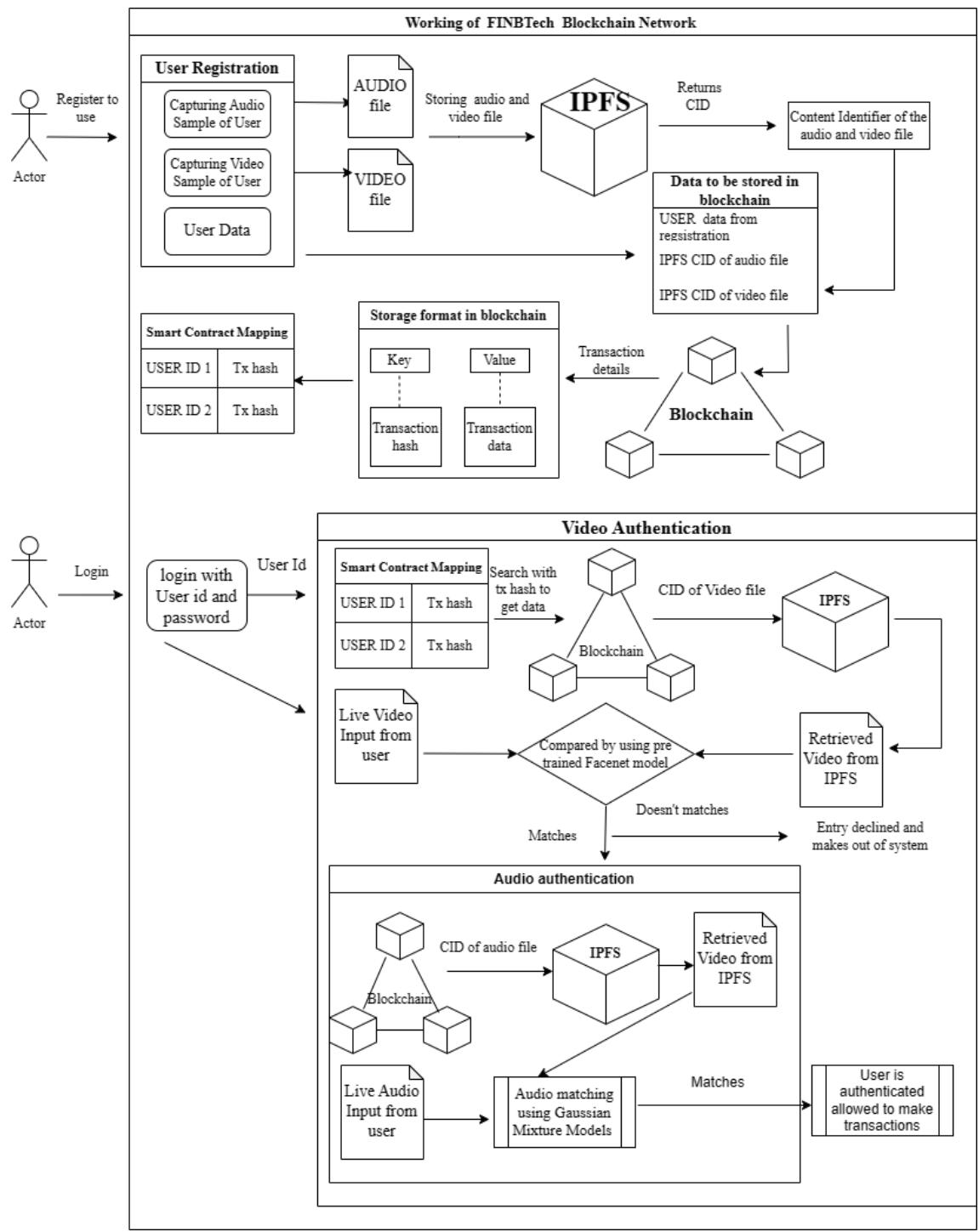

**Fig 2 :-  Storage and Retrieval Architecture of Proposed System**

### 3.1.4 User Id-Transaction Hash Mapping

The smart contract serves as a bridge between user IDs and their associated transaction hashes, establishing a clear and efficient connection within the blockchain. By linking user IDs to transaction data, users can swiftly locate and access their transactions without the need for extensive blockchain analysis, streamlining the process of searching for specific transaction records on the blockchain.

---

**Function TransactionHashMapping**

```
// Declare the smart contract
// Declare a mapping to associate userID with transactionHash
    mapping(string => bytes32) public userToTransaction;

    // Function to store the mapping
    function storeUserTransaction(string userID, bytes32 transactionHash) public {
        // Store the transactionHash for the given userID
        userToTransaction[userID] = transactionHash;
    }

    // Function to retrieve the transactionHash for a given userID
    function getTransactionHash(string userID) public view returns (bytes32) {
        // Retrieve and return the transactionHash associated with the userID
        return userToTransaction[userID];
    }
}
```

---

**3.2 Equitable Consensus Mechanism - Proof of Fair Chance (PoFC) algorithm**

PoFC is a distributed consensus algorithm designed to ensure that all participants in a blockchain network have a fair chance to participate in the consensus process. Designing a consensus algorithm with multiple constraints requires a careful balance between security, decentralization, and practical considerations. The following are constraints used in the Proof of Fair Chance algorithm.

1. Minimum computing power: To participate in the consensus process, nodes must have computing power that meets a minimum threshold. This helps to ensure that nodes are contributing enough computational resources to the network to help secure the blockchain. Let P be the computational power of a participant and T be the threshold value for computational power.

2. Minimum balance: To participate in the consensus process, nodes must also have a minimum balance of cryptocurrency. This helps to ensure that nodes have a stake in the network and are incentivized to act honestly. Let B be the cryptocurrency balance of a participant and B_min be the minimum balance required.

3. Maximum number of previously mined blocks: To prevent any one node from dominating the consensus process, nodes are limited in the number of blocks they can mine consecutively. This helps to ensure that the blockchain remains decentralized and resistant to manipulation. Let C be the consecutive blocks produced by a participant and C_max be the maximum consecutive blocks allowed.

4. Maximum network bandwidth: Nodes are limited in the amount of data they can send and receive over the network. This helps to ensure that the blockchain remains accessible to all nodes, even those with slower network connections. Let N be the network bandwidth of a participant and N_max be the maximum network bandwidth allowed.

5. Maximum storage capacity: Nodes are limited in the amount of storage they can allocate to the blockchain. This helps to ensure that the blockchain remains lightweight and accessible to all nodes. Let S be the storage capacity of a participant and S_max be the maximum storage capacity allowed.

The overall PoFC consensus can be expressed as the logical conjunction (AND) of all conditions:

PoFC = (P ≥ T) ∧ (B ≥ B_min) ∧ (C ≤ C_max) ∧ (N ≤ N_max) ∧ (S ≤ S_max)    ---- *(4)*

The equation (1) represents the chance the miner will get to mine a new block. The mining of new blocks will be made on the basis of a new consensus algorithm called Proof of Fair Chance. Proof of fair chance provides equal participation to all the miners. It also reduces difficult hash calculation. Proof of Fair chance is considered to be a better consensus algorithm because of its fairness and less computation along with equal participation.

### 3.3 User Voice and Video Authentication

User voice and video authentication is a security process that verifies a user's identity by analyzing their voice and video input, enhancing authentication and access control for systems and services. It provides an additional layer of security beyond traditional username and password methods.

### 3.3.1 User Login

When accessing the FinBTech blockchain network, users must provide their user ID and live video input for initial authentication, followed by a voice authentication step involving reading a provided paraphrase; successful completion of both steps grants access to the system.

### 3.3.2 Transaction Hash retrieval

The User ID-TransactionHash mapping smart contract is used to retrieve the transaction hash associated with the user who logs into the system, allowing efficient access to their transaction history on the blockchain.

| **Function RetrieveTransactionHash** |
| --- |
| // Function to retrieve the transactionHash for a given userID<br>function getTransactionHash(string userID) public view returns (bytes32)<br>   // Retrieve and return the transactionHash associated with the userID<br>   return userToTransaction[userID]; |

### 3.3.3 Blockchain Retrieval

To retrieve a transaction using a transaction hash from the TransactionStorage smart contract, you can directly call the transactions mapping with the transaction hash as the key.

| **Function getTransaction(transactionHash)** |
|---|
| // Access the transactions mapping using the transactionHash as the key<br>transaction = transactions[transactionHash]<br>// Return the transaction data<br>return transaction |

### 3.3.4 IPFS VideoRetrieval

The CID is obtained from the user's blockchain transaction, which serves as a unique reference to the video file's location in IPFS. The CID is utilized to fetch and retrieve the video file from IPFS, allowing access to the registered user's video data during login.

| **Function retrieveFilesFromIPFS(videoCID)** |
|---|
| // Retrieve the files from IPFS using the videoCID<br>retrievedFiles = ipfsClient.get(videoCID) |

### 3.3.5 Video Authentication

FaceNet512 generates the embedding vector from a face image as its input.FaceNet uses a face image as its input and produces a vector of 512 integers that represent the most crucial facial characteristics. This vector is known as an embedding in machine learning. FaceNet512 basically takes a face and turns it into a vector of 512 values. Ideally, similar faces also have similar embeddings. Facenet512 Maps high-dimensional data (like images) into low-dimensional representations (embeddings). Vectors can be understood as points in the Cartesian coordinate system, and embeddings are vectors. So, using its embeddings, we may plot an image of a face in the coordinate system.

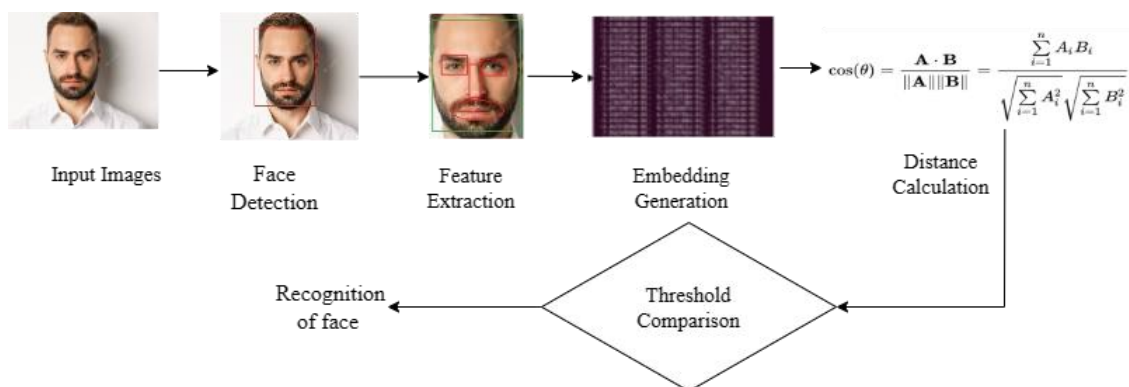

**Fig 3 :- Working of Face recognition using facenet**

FaceNet512 is a facial recognition model that processes facial images and extracts high-dimensional embeddings (vectors) representing unique facial features. The steps involved in the FaceNet512 process are shown in Figure 4 .

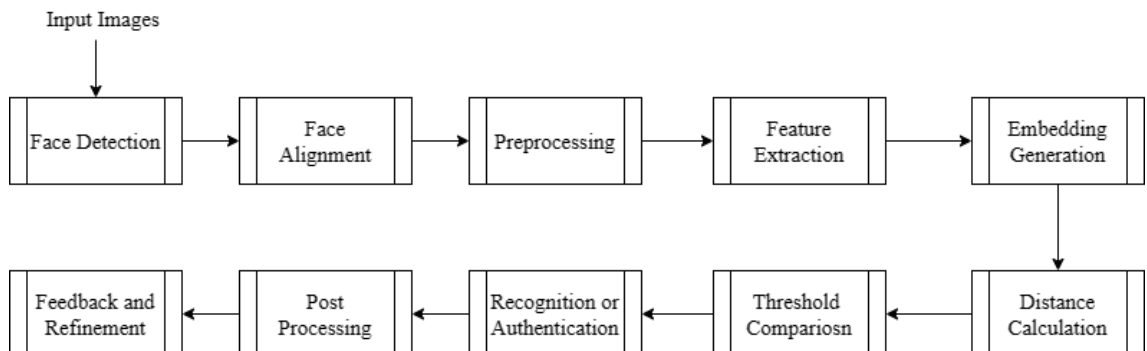

**Fig 4 :- Steps involved in face recognition using Facenet512**

### 3.3.5.1 Face Detection

Multi-task Cascaded Convolutional Neural Networks (MTCNN) is a deep learning-based model that can detect faces, facial landmarks, and perform face alignment in a single pass. It consists of three stages, each responsible for a specific task: face detection, facial landmark localization, and bounding box refinement. The steps involved in MTCNN are as follows:

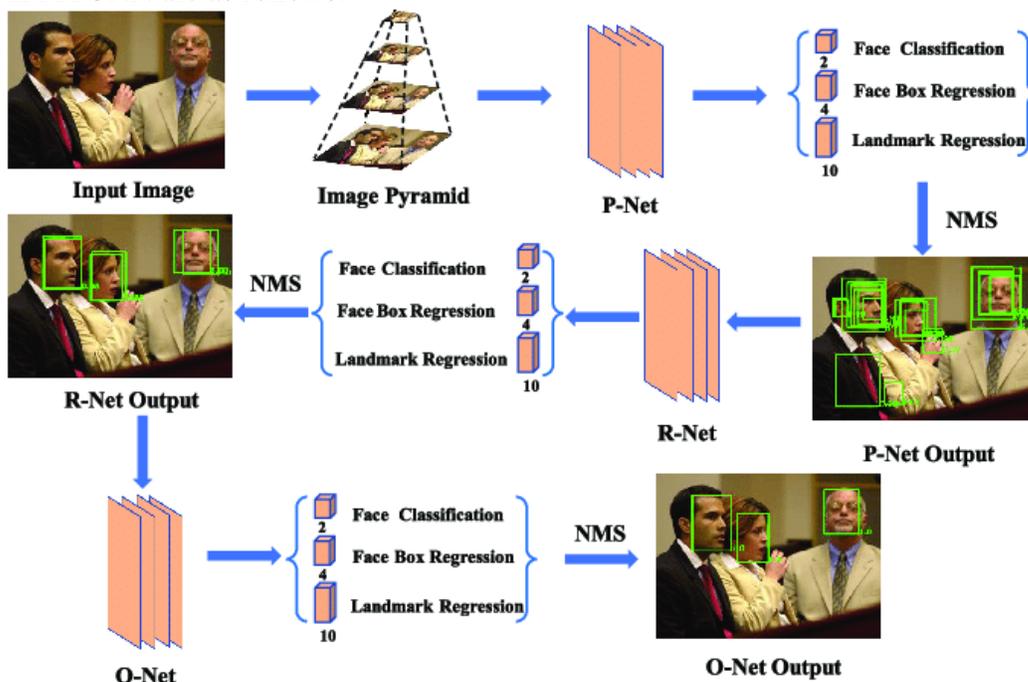

**Fig 5 :- Steps involved in face recognition using MTCNN**

### Stage 1: Proposal Network (P-Net)

**Image Pyramid Creation:** The input image in the video frame is denoted as ⬜. The scaling factor is denoted as ⬜, which determines how much the image to be scaled at each level of the pyramid. A typical value for scaling factor is 0.5, which reduces the image size to half at each level. The image is scaled down iteratively by the defined scaling factor. Each iteration produces a smaller version of the image, denoted as $I_n$. The scaling operation can be represented as:

Width of $I_n$ : $W_n = \alpha \cdot W_{n-1}$

Height of $I_n$ : $H_n = \alpha \cdot H_{n-1}$

where $W_n$ and $H_n$ represent the width and height of image $I_n$ respectively and $W_{n-1}$ and $H_{n-1}$ represent the width and height of the previous image $I_{n-1}$. The scaling factor α determines the reduction in size from one level of the pyramid to the next. For each scaled image $I_n$, run the P-Net to detect objects at that scale. The P-Net will generate a set of bounding boxes for potential objects in $I_n$.

**Convolutional Feature Extraction:** The P-Net applies a series of convolutional layers to extract feature maps from each scale of the image pyramid. The input image is passed through a stack of convolutional layers in the P-Net. These layers consist of learnable filters that slide over the input image to perform convolutions. Each convolution operation computes a weighted sum of the pixel values within a local receptive field. The result of each convolution is referred to as a feature map. The feature map at position $(x, y)$ in layer $L$ can be calculated as follows:

$F_L(x, y) = \sum_{i=1}^{H} \sum_{j=1}^{W} (I(x+i, y+j) \cdot K_L(i, j))$

where $F_L(x, y)$ is the value of the feature map at position $(x, y)$ in layer $L$.

$I(x+i, y+j)$ represents the pixel value at position $(x+i, y+j)$ in the input image.

$K_L(i,j)$ is the value of the kernel at position $(i,j)$ in layer $L$.

$H$ and $W$ represent the height and width of the kernel, respectively.

After each convolution operation, an activation function like Rectified Linear Unit (ReLU) is typically applied element-wise to the feature map. This introduces non-linearity into the network and helps capture complex patterns. The output of the activation function is the activated feature map $A_L(x, y) = \max(0, F_L(x, y))$. The P-Net typically consists of multiple convolutional layers. Each layer processes the feature maps from the previous layer to capture hierarchical features.

The final result of convolutional feature extraction is a set of feature maps that represent various patterns and details in the input image.

**Face Candidate Detection:** A sliding window approach is used on the feature maps obtained from the convolutional layers. This involves moving a small window across the feature map, examining different regions of the image at each position. At each position of the sliding window, the P-Net predicts a set of convolutional anchor boxes known as bounding boxes or region proposals. These anchor boxes come in different scales and aspect ratios to cover a wide range of object sizes and shapes.

For each anchor box, the P-Net predicts adjustments to the coordinates of the anchor box to better fit the faces within it. These adjustments are obtained through a set of convolutional layers followed by fully connected layers.

The adjusted bounding box coordinates $(x', y', w', h')$ for an anchor box at position $(x, y)$ in the feature map with coordinates $(x_a, y_a, w_a, h_a)$ can be calculated as follows:

$x' = x_a + \Delta x \cdot w_a$

$y' = y_a + \Delta y \cdot h_a$

$w' = w_a \cdot e^{\Delta w}$

$$h' = h_\square \cdot \square^{\Delta h}$$

where

$\Delta\square$ and $\Delta\square$ are the predicted adjustments to the x and y coordinates of the anchor box.

$\Delta\square$ and $\Delta h$ are the predicted adjustments to the width and height of the anchor box.

$\square$ is the base of the natural logarithm, approximately 2.71828.

For each anchor box, the P-Net predicts the probability ($\square$) of containing a face. This probability is often passed through a sigmoid activation function to produce a probability score indicating the likelihood of containing a face.

These equations and processes are applied to each anchor box generated at different positions on the feature map, resulting in a set of candidate bounding boxes for potential faces in the input image. These candidates are further refined and filtered in subsequent stages of the MTCNN framework.

**Non-Maximum Suppression (NMS):** Overlapping bounding boxes with lower confidence scores are removed using non-maximum suppression, leaving a set of highly probable face candidates. Each bounding box is associated with a confidence score indicating the likelihood of it containing an object. The confidence score of bounding box $\square$ be denoted as $\square_\square$.

The bounding boxes are sorted in descending order based on their confidence scores. The bounding box with the highest confidence score will be considered first. Starting with the bounding box that has the highest confidence score, the sorted list of bounding boxes are iterated. For each box in the iteration, calculate the intersection-over-union (IoU) with the previously selected boxes. The IoU between two bounding boxes, $\square_1$ and $\square_2$ can be calculated as follows:

IoU($\square_1,\square_2$)=Area of Overlap/Area of Union

If the IoU between $\square_\square$ , the currently considered box and any previously selected box ($\square_\square$) is above the threshold of 0.5, it means that the two boxes significantly overlap, and one of them should be suppressed. To decide which box to keep, compare their confidence scores. The box with the higher confidence score is retained, and the other is suppressed. After completing the NMS iteration, a set of non-overlapping bounding boxes are left that represent the final candidate detections for faces in the input image.

NMS(boxes,confidence_scores,IoU_threshold)=selected_boxes

Where boxes is the list of candidate bounding boxes.

confidence_scores is the corresponding list of confidence scores.

IoU_threshold is the threshold above which bounding boxes are considered to significantly overlap.

selected_boxes is the final list of selected, non-overlapping bounding boxes.

## Stage 2: Refinement Network (R-Net)

**Image Crop:** For each candidate bounding box, extract the region of the original image that corresponds to that bounding box. This cropped region contains face and its surrounding context. The cropped region is denoted as $\square_{crop}$. The process of cropping an image region can be represented as follows:

$\square_{crop}=\square(\square,\square,\square,h)$

where I represents the original input image, and $\square,\square,\square$ and $h$ represent the coordinates and dimensions of the candidate bounding box. The values of $\square,\square,\square$ and $h$ are determined based on the location and size of the bounding box. □ and □ represent the top-left corner coordinates of the bounding box, while □ and h represent the width and height of the bounding box, respectively.

The cropped regions $□_{crop}$ are used as inputs to the Refinement Network. The Refinement Network further processes these regions to refine the object's localization and improve the quality of the detection.

**Resizing and Preprocessing:** The cropped regions are resized to a fixed size and preprocessed before being fed into the R-Net.

To ensure that all cropped regions have a consistent size and aspect ratio for input into the R-Net, each cropped region is resized to a fixed size. This resizing ensures that the R-Net can process regions of interest with uniform dimensions. The resized regions be denoted as $□_{resized}$.

The process of resizing an image region can be represented as follows:

$$□_{resized} = resize(□_{crop}, target\_size)$$

Where $□_{resized}$ represents the resized region of interest.

$□_{crop}$ represents the cropped region from the P-Net.

resize is a function that resizes the image $□_{crop}$ to the specified target_size. The target size is typically a fixed dimension.

Apply any necessary preprocessing steps to the resized regions before feeding them into the R-Net. Common preprocessing steps include mean subtraction, scaling pixel values to a range [0, 1] and channel-wise normalization.

**Convolutional Feature Extraction:** Convolutional layer is applied in R-net to extract feature maps from the face candidates.

Convolutional Feature Extraction in the R-Net involves passing the preprocessed and resized image regions through a series of convolutional layers to extract relevant features. These features are used to refine the localization and attributes of objects within the regions.

The resized regions ($□_{resized}$) are passed through a stack of convolutional layers in the R-Net. These convolutional layers consist of learnable filters or kernels that slide over the input regions to perform convolutions. Each convolution operation computes a weighted sum of the pixel values within a local receptive field. The result of each convolution is referred to as a feature map.

After each convolution operation, an activation function, ReLU, is applied element-wise to the feature maps. This introduces non-linearity into the network and helps capture complex patterns. The output of the activation function is the activated feature map ($□_□(□,□)$).

The equation for the activated feature map can be represented as:

$$□_□(□,□) = max(0, □_□(□,□))$$

where $□_□(□,□)$ is the value of the feature map at position ($□,□$) in layer □ and $□_□(□,□)$ is the activated feature map at the same position.

The R-Net typically consists of multiple convolutional layers. Each layer processes the feature maps from the previous layer to capture hierarchical features. The convolution operation for a single feature map can be represented as follows:

$$□_□(□,□) = \sum_{□=1}\sum_{□=1}□(□_{resized}(□+□,□+□)\cdot □_□(□,□))$$

Where

$\square_\square(\square,\square)$ is the value of the feature map at position $(\square,\square)$ in layer $\square$.

$\square_{\text{resized}}(\square+\square,\square+\square)$ represents the pixel value at position $(\square+\square,\square+\square)$ in the resized input region.

$\square_\square(\square,\square)$ is the value of the kernel at position $(\square,\square)$ in layer L.

$\square$ and $\square$ represent the height and width of the kernel, respectively.

The extracted features in the form of activated feature maps ($\square_\square(\square,\square)$) capture relevant patterns and details within the resized regions, allowing the R-Net to refine object localization and attributes, such as accurately detecting and refining the position and attributes of a face within the resized region.

**Face Classification:** The R-Net classifies each face candidate as a face or non-face based on the features extracted. It also provides bounding box regression to refine the bounding box coordinates.

The R-Net takes as input the Region of Interests (ROIs) that have been refined and preprocessed in previous stages, such as the P-Net and feature extraction. The ROIs are passed through a series of convolutional layers in the R-Net. These layers are responsible for learning hierarchical features from the input ROIs.

After convolutional layers, an activation function is applied element-wise to the resulting feature maps. The feature maps are flattened and fed into one or more fully connected layers. These layers perform the actual classification task. The output of these fully connected layers is a single value, often referred to as the logit ($\square$), which represents the network's confidence regarding the presence of a face.

To obtain a probability score indicating the likelihood of a face being present, a sigmoid activation function is applied to the logit $\square$. The sigmoid function maps the logit to a value between 0 and 1:

$$\square = 1/1 + \square^{-\square}$$

Where: $\square$ is the probability that the ROI contains a face.

$\square$ is the logit obtained from the fully connected layers.

In this equation, $\square$ represents the network's confidence in the presence of a face within the ROI. If $\square$ is above a certain threshold of 0.5, the ROI is classified as containing a face; otherwise, it is classified as not containing a face.

**Non-maximum suppression (NMS):** NMS is applied again to eliminate redundant bounding boxes, leaving a set of refined face candidates. The process of non-maximum suppression can be represented mathematically as follows:

$$NMS(\square,\square,IoU\_threshold) = selected\_boxes$$

Where $\square$ represents the list of bounding boxes.

$\square$ represents the list of confidence scores associated with each bounding box.

IoU_threshold is the threshold above which bounding boxes are considered to significantly overlap.

selected_boxes is the final list of selected, non-overlapping bounding boxes.

**Stage 3: Output Network (O-Net)**

**Image Crop:** Similar to Stage 2, the refined face candidates from the R-Net are cropped from the original image.

**Resizing and Preprocessing:** The cropped regions are resized and preprocessed for input to the O-Net.

**Convolutional Feature Extraction:** The O-Net further processes the face candidates with convolutional layers to extract feature maps.

**Facial Landmark Localization:** The O-Net is designed to not only classify faces but also locate facial landmarks, such as eyes, nose, and mouth. It outputs landmark coordinates for each face candidate.

**Face Classification and Bounding Box Refinement:** Similar to the R-Net, the O-Net performs face classification and bounding box regression for precise face localization.

**NMS:** Non-maximum suppression is applied one final time to obtain the most accurate and non-overlapping face bounding boxes along with landmark coordinates.

The final output of MTCNN consists of detected faces' bounding boxes and facial landmark positions.

### 3.3.5.2 Face Alignment

Once faces are detected, the system may perform face alignment to normalize the position and scale of the detected faces. This step ensures that facial features are consistently located in the same positions, improving the accuracy of the recognition process.

Face alignment involves transforming facial landmarks in an input face image to a canonical or standardized configuration. This process often includes translation, rotation, and scaling operations to ensure that key facial features are consistently positioned and oriented.

The input face image represented as $\square$. The facial landmarks N are detected in the input image. These landmarks are typically represented as a set of coordinates $(\square_\square, \square_\square)$ for $\square=1,2,\ldots,\square$. These landmarks define the positions of key facial features such as eyes, nose, and mouth.

The canonical configuration is denoted as $\square$, which represents the desired positions and orientations of the facial landmarks in a standardized manner. The transformation is calculated that aligns the detected landmarks to the canonical configuration. This transformation often includes translation, rotation, and scaling. The transformation can be represented as a combination of the following operations:

**Translation:** Shifting the landmarks by $(\Delta\square, \Delta\square)$ to align them with the canonical positions:

$$\square_\square' = \square_\square + \Delta\square$$
$$\square_\square' = \square_\square + \Delta\square$$

**Rotation:** Rotating the landmarks by an angle $\square$ to match the canonical orientation:

$$\square_\square'' = \square_\square' \cos(\square) - \square_\square' \sin(\square)$$
$$\square_\square'' = \square_\square' \sin(\square) + \square_\square' \cos(\square)$$

**Scaling:** Scaling the landmarks to match the canonical distances between key features like eye-to-eye distance:

$$\square_\square''' = \square \cdot \square_\square''$$
$$\square_\square''' = \square \cdot \square_\square''$$

Apply the alignment transformation to the entire input image $\square$ to obtain the aligned or warped face image aligned $I_{aligned}$. This is done using interpolation techniques:

$$\square_{aligned}(\square,\square) = \square(\square_\square''', \square_\square''')$$

Here $\square_\square'''$ and $\square_\square'''$ are the transformed coordinates of each pixel $(\square,\square)$ in the output image. The resulting $\square_{aligned}$ is an aligned face image where facial landmarks are consistent with the canonical configuration.

### 3.3.5.3 Preprocessing

The facial images are preprocessed to enhance their quality and reduce noise. Preprocessing in FaceNet-512 involves data preparation steps to ensure that input face images are in a suitable format for accurate face recognition.

Input face images are resized to a fixed resolution of 160x160 pixels. This ensures that all images have the same dimensions for consistency during processing. The pixel values in the resized images are normalized to a specific range [0, 1] or [-1, 1]. Normalization helps in reducing the impact of variations in lighting conditions. The faces are detected in the input images, and align them to a canonical position. Face alignment techniques ensure that the faces are centered and oriented consistently for recognition.

The random transformations such as cropping, rotation, and horizontal flipping are applied to increase the diversity of the training data. Data augmentation helps the model generalize better to different variations.

Prewhitening is applied to the image data to remove variations in brightness and contrast. It involves centering the data around zero mean and unit variance by subtracting the mean pixel value and dividing by the standard deviation. The preprocessed face images are organized into batches before feeding them into the neural network for inference. Batching enables parallel processing and helps improve computational efficiency.

The preprocessed face images are converted into tensors, which are multi-dimensional arrays suitable for deep learning frameworks. Tensors allow efficient processing by the neural network.

### 3.3.5.4 Feature Extraction

FaceNet512 employs a deep convolutional neural network to extract high-level features from the preprocessed facial images. The network architecture typically consists of multiple convolutional layers, followed by fully connected layers. In the case of FaceNet512, it produces 512-dimensional feature embeddings for each face.

The input data consists of preprocessed face images that have been resized, normalized, aligned, and prewhitened. Feature extraction is performed by a Convolutional Neural Network architecture, which is a deep neural network designed to automatically learn hierarchical and abstract features from images.

The CNN consists of multiple layers, including convolutional layers and pooling layers. Convolutional layers apply learnable filters to the input image, which slide across the image to detect patterns and features. Each convolutional layer captures different levels of abstraction, from simple edges and textures to complex facial features.

After each convolutional layer, non-linear activation functions like ReLU are applied element-wise to introduce non-linearity and enhance feature representation. Pooling layers downsample feature maps, reducing their spatial dimensions while retaining important information. This helps in reducing computational complexity.

The output of the CNN's convolutional and pooling layers is often flattened and passed through fully connected layers. These fully connected layers perform high-level feature extraction and dimension reduction. The last fully connected layer, often called the "embedding layer," produces a fixed-length feature vector for each input face image. The feature vector encodes facial features and characteristics that are discriminative for face recognition.

Feature vectors are typically L2 normalized to make them invariant to scaling. This involves dividing each feature vector by its L2 norm. The resulting normalized feature vectors reside in a high-dimensional feature space where the similarity between vectors corresponds to facial similarity. Cosine similarity is often used to measure similarity in this space.

### 3.3.5.5 Embedding Generation

Embedding generation in FaceNet-512 refers to the process of producing a high-dimensional feature vector from an input face image. These embeddings are designed to be highly discriminative, allowing for accurate face recognition based on similarity measurements in the feature space.

The input to the embedding generation process is a preprocessed face image, which has undergone resizing, normalization, alignment, and possibly prewhitening.

**CNN Operation:** Embedding generation is carried out by a deep CNN architecture. The CNN is composed of multiple layers, including convolutional layers, pooling layers, activation functions, and fully connected layers.

The convolution operation between an input feature map $\square$ and a convolutional filter $\square$ is represented as:

$$(\square * \square)(\square,\square)=\sum_\square \sum_\square \square_\square(\square+\square,\square+\square)\cdot \ \square(\square,\square)$$

where $\square$ and $\square$ are the spatial coordinates and $(\square,\square)$ are the coordinates within the filter $\square$.

**Feature Extraction:** The CNN processes the input image through its layers to extract hierarchical and discriminative features. Each convolutional layer applies learnable filters to detect patterns and features in the image, while pooling layers downsample and retain important information.

ReLU is a commonly used activation function. It is defined as ReLU($\square$)=max(0,$\square$). It introduces non-linearity by setting all negative values to zero.

Max-pooling downsampled the feature map by selecting the maximum value in each pooling window. For a window size of 2×2, the operation is:

Max-Pooling($\square$)($\square,\square$)=max($\square$(2$\square$,2$\square$),$\square$(2$\square$,2$\square$+1),$\square$(2$\square$+1,2$\square$),$\square$(2$\square$+1,2$\square$+1))

where $\square$ and $\square$ are the spatial coordinates.

**Fully Connected Layers:** After the convolutional and pooling layers, the feature maps are typically flattened and passed through one or more fully connected layers. These fully connected layers perform high-level feature extraction and dimension reduction.

In a fully connected layer, a linear transformation is applied to the flattened feature map $\square$ with learned weights $\square$ and biases $\square$, followed by an activation function:

$$\square=\square \cdot \ \square+\square$$

The output $\square$ represents the logits before the activation function.

**Embedding Layer:** The last fully connected layer, often referred to as the "embedding layer," produces the final embedding vector for the input image. This embedding vector is a high-dimensional feature representation of the input face image, capturing its unique characteristics.

**L2 Normalization:** The embedding vector is L2 normalized, which involves dividing each element of the vector by its L2 norm. L2 normalization ensures that the embedding vector has a fixed length, making it invariant to scaling.

**Feature Vector:** The normalized embedding vector is now a feature vector that resides in a high-dimensional feature space. The elements of this vector encode the distinctive features of the face image, making it suitable for recognition.

**Similarity Measurement:** To perform face recognition, the similarity between embedding vectors is measured using a metric like cosine similarity. A smaller cosine distance between two embeddings indicates higher similarity between the corresponding face images.

**L2 Normalization:** L2 normalization scales the embedding vector $E$ to have unit length:

$$E_{normalized} = \frac{E}{\|E\|_2}$$

Where $\|E\|_2$ is the L2 norm of $E$, calculated as the square root of the sum of squares of its elements.

### 3.3.5.6 Distance Calculation

To compare two facial embeddings and determine if they belong to the same person, a distance metric is computed between them.

In FaceNet, distance calculation is a crucial step in face recognition. It involves measuring the similarity between two face embeddings to determine whether they represent the same person or different people. Cosine similarity is commonly used for this purpose.

Cosine similarity measures the cosine of the angle between two vectors. Consider there are two face embeddings $E_1$ and $E_2$ each representing a face:

$$E_1 = [E_{1,1}, E_{1,2}, \ldots, E_{1,n}]$$
$$E_2 = [E_{2,1}, E_{2,2}, \ldots, E_{2,n}]$$

Where $n$ is the dimensionality of the embeddings, and $E_{1,n}$ and $E_{2,n}$ are the components of the embeddings.

The cosine similarity $\cos(\theta)$ between these embeddings is calculated as follows:

$$\text{Cosine Similarity}(E_1, E_2) = \frac{E_1 \cdot E_2}{\|E_1\|_2 \cdot \|E_2\|_2}$$

where $E_1 \cdot E_2$ represents the dot product between the two embeddings.

$\|E_1\|_2$ and $\|E_2\|_2$ are the Euclidean norms of the embeddings, calculated as the square root of the sum of squares of their components.

The result of the cosine similarity calculation is a value in the range of -1 to 1. If the two embeddings are very similar and represent the same person, the cosine similarity will be close to 1. If the embeddings are dissimilar and represent different people, the cosine similarity will be close to -1. If the embeddings are somewhat similar but not identical, the cosine similarity will be between -1 and 1, with values closer to 1 indicating higher similarity.

### 3.3.5.7 Threshold Comparison

The calculated distance is compared against a predefined threshold. If the distance is below the threshold, the two facial embeddings are considered a match, suggesting that the faces belong to the same person. If the distance exceeds the threshold, they are considered different.

The predefined threshold value is denoted as □. The computed cosine similarity is compared with the predefined threshold as follows:

Decision=   Same Person            if Cosine Similarity(□1,□2)≥□ threshold

Different People        if Cosine Similarity(□1,□2)<□threshold

If the cosine similarity is greater than or equal to the threshold, it is considered a positive match, indicating that the two embeddings likely represent the same person. If the cosine similarity is less than the threshold, it is considered a negative match, suggesting that the embeddings likely represent different individuals.

### 3.3.6 Voice Authentication

After the video of the user is authenticated, the user reads the paraphrase, and their voice is recorded. Voice authentication is performed using GMM, comparing the recorded voice during the login with the voice input provided by the user during registration. Access is granted if the comparison exceeds the threshold value.

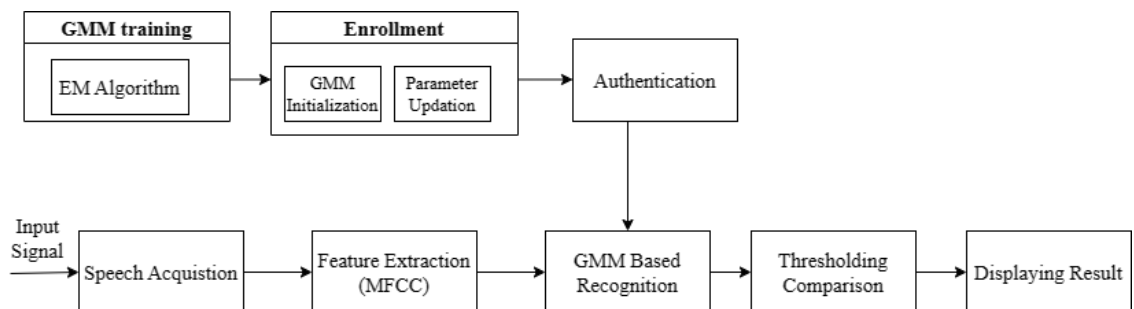

**Fig6 :- Steps involved in Voice Authentication using GMM**

### 3.3.6.1 Feature extraction:
Feature extraction transforms raw audio data into a set of informative features.

**Mel-Frequency Cepstral Coefficients(MFCC):** The key steps in computing MFCCs involve the Discrete Fourier Transform (DFT), mel filtering, and cepstral coefficient computation. In feature extraction, the audio signals are divided into short overlapping frames. Each frame is multiplied by a Hamming window. The DFT is computed for the windowed frame. The power spectrum is computed by taking the magnitude of the DFT coefficients squared. The power spectrum is mapped onto the mel scale using a set of triangular filters. The natural logarithm of the filtered energies is taken. Discrete Cosine Transform (DCT) is applied to the log-filterbank energies to obtain the MFCCs.

The MFCCs can be represented as a vector: MFCC =$[□_1, □_2, \ldots, □_□]$

where □ is the number of coefficients.

**Pitch Features:** Pitch features capture information about the fundamental frequency called pitch of the voice signal. One common measure is the pitch period (□):

□=argmax$_□$□(□)

where □(□) represents the autocorrelation function, and □ is the time delay.

**Formant Frequencies:** Formant frequencies represent the resonant frequencies in the vocal tract and are important for characterizing vowel sounds. The first two formants ($\square 1$ and $\square 2$) can be estimated using Linear Predictive Coding (LPC) analysis:

$$\square 1 = \frac{c}{4L1}$$
$$\square 2 = \frac{c}{4L2}$$

Where $\square$ is the speed of sound, and $\square 1$ and $\square 2$ are the lengths of the vocal tract resonators.

**Spectral Features:** Spectral features such as spectral centroid ($\square\square$), spectral bandwidth ($\square\square$), and spectral rolloff ($\square\square$) can be computed as follows:

$$SC = \frac{\sum_{k=0}^{N-1} f(k)X(k)}{\sum_{k=0}^{N-1} X(k)}$$

$$SB = \sqrt{\frac{\sum_{k=0}^{N-1} (f(k) - SC)^2 X(k)}{\sum_{k=0}^{N-1} X(k)}}$$

$$SC = \frac{\sum_{k=0}^{N-1} X(k)}{2}$$

Where $\square(\square)$ represents the frequency of bin $\square$ in the DFT, $\square(\square)$ is the magnitude of the DFT coefficient at bin $\square$, $\square$ is the number of bins.

**Delta and Delta-Delta Features**: Delta and delta-delta features capture the temporal dynamics of the static features. These can be represented as the first and second derivatives of the static features, respectively:

$$\Delta c_n(t) = c_n(t+1) - c_n(t-1)$$
$$\Delta\Delta c_n(t) = c_n(t+1) - c_n(t-1)$$

Where $c_n(t)$ represents the n-th coefficient at time $\square$.

### 3.3.6.2 Training

Training involves estimating the model's parameters including means, variances, and mixture weights of the Gaussian components. The training process aims to fit the GMM to a given dataset in a way that it can capture the underlying data distribution.

Consider a dataset $\square = \{\square_1, \square_2, \dots, \square_N\}$ where $x_i$ represents a data point. To fit a GMM with $\square$ Gaussian components, the following steps are followed:

**Initialization:** The following parameters of the GMM are initialized:

- Mixture weights: $\square_\square$ for $\square = 1, 2, \dots, \square$ where $\sum_{k=1}^{K} \square_\square = 1$
- Gaussian means: $\square_\square$ for $k = 1, 2, \dots, K$.
- Gaussian variances: $\square_\square{}^2$ for $\square = 1, 2, \dots, K$.

**Expectation-Maximization (EM) Algorithm:** EM algorithm involves two main steps: the Expectation-step (E-step) and the Maximization-step (M-step). These steps are iteratively performed until convergence.

**E-step():** The posterior probabilities or responsibilities of each data point $\square_\square$ belonging to each Gaussian component $\square$ is calculated as follows:

$$\square_{\square\square} = \frac{\pi_k N(x_i \mid \mu_k, \sigma_k^2)}{\sum_{k=1}^{K} \pi \; \pi_k N(x_i \mid \mu_k, \sigma_k^2)}$$

**M-step ():** The GMM parameters are updated based on the responsibilities calculated in the E-step. The mixture weights are updated as follows:

$$\pi_k = \frac{1}{N} \sum_{i=1}^{N} w_{ik}$$

The Gaussian means are updated as follows:

$$\mu_k = \frac{\sum_{i=1}^{N} w_{ik} \cdot x_i}{\sum_{i=1}^{N} w_{ik}}$$

The Gaussian variances are updated as follows::

$$\sigma_k^2 = \frac{\sum_{i=1}^{N} w_{ik} \cdot (x_i - \mu_k)^2}{\sum_{i=1}^{N} w_{ik}}$$

**Normalization:** The mixture weights has still sum up to 1:$\sum_{k=1}^{K} \square_\square = 1$

**Convergence:** The E-step and M-step are repeated until the GMM parameters converge.

**Model Selection:** The number of Gaussian components in the GMM is a critical parameter. Model selection techniques, such as the Akaike Information Criterion (AIC) or the Bayesian Information Criterion (BIC), can be used to determine the optimal number of components for a given dataset.

**Final GMM:** Once the training process is complete and the GMM has converged, GMM is trained with $\square$ Gaussian components, and the parameters $\pi_k, \mu_k \ and \ \sigma_k^2$ for each component are estimated to best fit the training data.

### 3.3.6.3 Enrollment

Enrollment involves the process of creating individual GMMs for each enrolled user or speaker. Each speaker's GMM is trained using their voice samples.

**Data Collection:** The voice samples are collected from the users who want to enroll in the voice authentication system. These voice samples should be representative of each user's speech. The relevant features are extracted from each voice sample.

**GMM Initialization:** For each enrolled user, an empty GMM is initialized with $\square$ Gaussian components. The choice of $\square$, the number of components, depends on the application and may be determined through model selection techniques. The GMM parameters are initialized for each component of each user's GMM as follows:

- Mixture weights: $\square_\square$ for $\square = 1,2,\ldots,\square$ where $\sum_{k=1}^{K} \square_\square = 1$
- Gaussian means: $\square_\square$ for $k=1,2,\ldots,K$.
- Gaussian variances: $\square_\square^2$ for $\square=1,2,\ldots,K$.

**Training User-Specific GMMs:** For each enrolled user $\square$, their voice samples and the corresponding extracted features $\square_\square$ are used to train their user-specific GMM. Expectation-Maximization (EM) algorithm is applied to estimate the GMM parameters for each user's GMM. The EM algorithm will iteratively update the mixture weights, means, and variances to maximize the likelihood of observing the user's voice data given the GMM.

**Updating Parameters:** After each iteration of the EM algorithm, the GMM parameters $(\pi_k, \mu_k \ and \ \sigma_k^2)$ are updated for each user $\square$ and each Gaussian component $\square$

**Convergence:** The training process continues until the GMM parameters converge, meaning that there is little or no change in the parameters between successive iterations.

**Enrollment Complete:** Once the enrollment process is complete, user-specific GMMs for each enrolled speaker is obtained. These GMMs are capable of modeling the characteristics of each speaker's voice.

**Authentication:** During the authentication phase, when a user provides their voice sample for verification, the features extracted from that sample are compared to the GMMs of the enrolled users. The likelihood scores calculated for each user's GMM can be used to determine whether the voice sample matches the enrolled user.

### 3.3.6.4 Authentication

Authentication involves comparing a voice sample provided by a user to the GMMs of enrolled users to determine whether the voice sample matches an enrolled user's voice.

Consider the enrollment phase having three users as User A, User B, and User C have provided voice samples and their respective GMMs have been trained. A new user, User X wants to be authenticated using their voice sample.

**Enrollment Phase**:
- **User A's GMM:**
  - Mixture weights: $\pi_{A1}, \pi_{A2}$
  - Gaussian means: $\mu_{A1}, \mu_{A2}$
  - Gaussian variances: $\sigma_{A1}^2, \sigma_{A2}^2$
- **User B's GMM:**
  - Mixture weights: $\pi_{B1}, \pi_{B2}$
  - Gaussian means: $\mu_{B1}, \mu_{B2}$
  - Gaussian variances: $\sigma_{B1}^2, \sigma_{B2}^2$
- **User C's GMM:**
  - Mixture weights: $\pi_{C1}, \pi_{C2}$
  - Gaussian means: $\mu_{C1}, \mu_{C2}$
  - Gaussian variances: $\sigma_{C1}^2, \sigma_{C2}^2$

**Authentication Phase**: User X provides a voice sample, and relevant features are extracted from it. Let's denote the features as $x_x$. The likelihood score for User X's voice sample is calculated with each enrolled user's GMM. The likelihood score represents how well the voice sample matches the model of each enrolled user. For each enrolled user $i$ (A, B, C), the likelihood score ($L_i$) is calculated as follows:

$$L_i = \sum_{j=1}^{M_x} log(\sum_{k=1}^{K_i} \pi_{ik}.N(x_j \mid \mu_{ik}, \sigma_{ik}^2))$$

$M_x$ is the number of feature vectors in User X's voice sample.

$K_i$ is the number of Gaussian components in User i's GMM.

$x_j$ represents the $j$-th feature vector in User X's voice sample.

$\pi_k, \mu_k \text{ and } \sigma_k^2$ are the GMM parameters for User i's GMM.

**Decision**: The likelihood scores for User X is compared with each enrolled user's GMM. The user with the highest likelihood score is considered the most likely match. If User X's likelihood score for User A is the highest, then User X is authenticated as User A. Optionally, you can set a threshold on the likelihood scores to make the authentication decision more robust. If none of the likelihood scores exceed the threshold, the authentication is denied.

**Authentication Result:**If User X is authenticated as one of the enrolled users, access or authentication is granted. If none of the enrolled users' likelihood scores surpass the threshold or if there is no clear match, access is denied.

### 3.3.6.5 Thresholding

Thresholding is the process of making a decision on whether to accept or reject a user's voice sample based on the likelihood scores calculated from the user's voice data and the enrolled user models. By setting an appropriate threshold, the level of confidence required for authentication can be controlled.

**Likelihood Scores:**Recall that during the authentication phase, you calculate likelihood scores ($\square\square$) for each enrolled user's GMM when comparing their models to the user's voice sample. These likelihood scores represent how well the voice sample matches each enrolled user's voice model:

- $\square_X$ is the number of feature vectors in the user's voice sample.
- $\square_\square$ is the number of Gaussian components in the GMM of enrolled user $\square$
- $\square_\square$ represents the $\square$-th feature vector in the user's voice sample.
- $\square_{\square\square}$, $\square_{\square\square}$, and $\square_{\square\square}{}^2$ are the GMM parameters for enrolled user $\square$'s GMM.

**Thresholding Decision:**To make an authentication decision, you compare the highest likelihood score ($\square_{max}$) obtained among all enrolled users with a predefined threshold ($\square$).

$$\square_{max}=\max_{\square}{\square_\square}$$

where $\square_{max}$ is the highest likelihood score among all enrolled users.

$\square$ is the threshold value.

**Authentication Decision:** If $\square_{max}$ is greater than the threshold $\square$, the user is authenticated, and access is granted. If $\square_{max}$ is less than or equal to the threshold $\square$, the authentication is denied, and access is not granted.

**Threshold Selection:**The choice of the threshold ($\square$) is crucial and depends on the desired security level and system performance. A higher threshold increases security but may result in more false rejections (legitimate users being denied access). Conversely, a lower threshold can reduce false rejections but may increase the risk of false acceptances (unauthorized users gaining access).

### 3.3.6.6 Adaptation

Adaptation in a Gaussian Mixture Model (GMM) for voice authentication refers to the process of updating an enrolled user's GMM model to account for changes in their voice characteristics over time. This is crucial to maintain the accuracy of the authentication system as a user's voice may change due to factors like aging or illness. One common technique for adaptation is Maximum Likelihood Linear Regression (MLLR). MLLR is a technique used to adapt the GMM parameters, specifically the mean vectors $\square_{\square\square}$ of the Gaussian components, to better match the current speaker's voice characteristics. The adaptation is typically done using adaptation data collected from the same speaker during the enrollment phase.

**Original GMM**: Let's denote the original GMM for an enrolled user i as $\Theta_i$ , with parameters including mixture weights ($\pi_i$ ), means ($\mu_i$), and variances ($\sigma_i{}^2$ ), where k represents the Gaussian component index.

**Adaptation Data**:During the enrollment phase, you collect adaptation data, denoted as $X_{adapt,i}$ ,which consists of feature vectors (e.g., MFCCs) from the same user's voice samples that represent their current voice characteristics.

**Compute Transformation Matrix:**Calculate a linear transformation matrix $W_i$ that minimizes the difference between the adapted GMM and the adaptation data. This is typically done by solving the following equation:

$$W_i = \frac{C_i}{S_i}$$

Where $S_i$ is the matrix of second-order statistics of the adaptation data and $C_i$ is the matrix of second-order statistics of the adapted GMM using the original parameters.

These matrices can be computed as follows:

$$S_i = \frac{1}{N} \sum_{j=1}^{N} x_{adapt,i,j} x_{adapt,i,j}^T$$

$$C_i = \sum_{k=1}^{K_i} \pi_{ik}\mu_{ik}\mu_{ik}^T + \sigma_{ik}^2 I$$

Where N is the number of frames in the adaptation data and $x_{adapt,i,j}$ represents the j-th feature vector in the adaptation data.

**Apply Transformation**:Apply the transformation matrix $W_i$ to adapt the mean vectors of the Gaussian components:

$$\mu_{ik}^{adapted} = W_i\mu_{ik}$$

This transformation aligns the GMM's mean vectors with the current voice characteristics of the user.

**Updated GMM**:The adapted GMM for user i ($\theta_i^{adapted}$) now has the adapted mean vectors while keeping the original mixture weights and variances:

$$\theta_i^{adapted} = \{\pi_i, \mu_{ik}^{adapted}, \sigma_{ik}^2\}$$

The adapted GMM can then be used for voice authentication, taking into account the user's current voice characteristics. This process allows the GMM to adapt to individual users over time, making the authentication system more robust to changes in their voices.

## 4. Results and Discussion

**Fig 7 :- Storing the file in IPFS**

**Fig 8:- Storing the file CID in the blockchain**

Figure 7 shows storing of audio and video files in IPFS and figure 8 shows storage of CID returned by IPFS and user details in Blockchain.

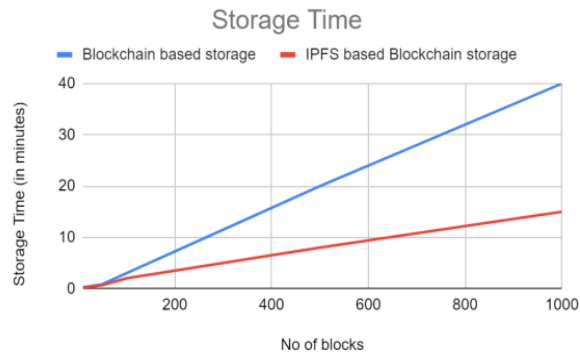

**Fig 9 :- Storage time for Blockchain and IPFS based Blockchain based on no of blocks**

Figure 9 shows the comparison of storage time needed for Blockchain Storage and IPFS Based Blockchain storage based on the number of blocks. The storage time needed for IPFS-based blockchain is less than the Blockchain-based storage. Therefore IPFS based blockchain storage takes minimum time to store a file even if there are more blocks.

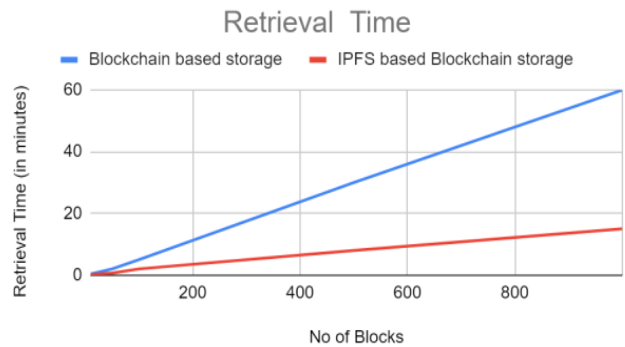

**Fig 10:- Retrieval time for Blockchain and IPFS based Blockchain based on no of blocks**

Figure 10 shows the comparison of retrieval time needed for Blockchain Storage and IPFS Based Blockchain storage based on the number of blocks. The time taken to retrieve a file from an IPFS-based blockchain is less than the Blockchain-based storage. Therefore IPFS based blockchain storage takes minimum time to retrieve a file even if the file size is large.

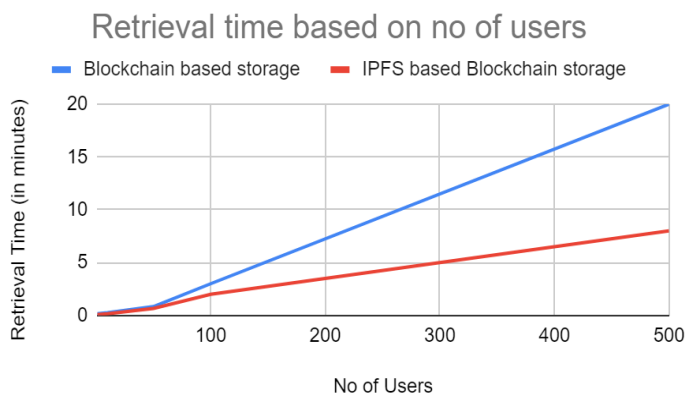

**Fig 11:- Retrieval time for Blockchain and IPFS based Blockchain based on the of users**

      Figure 11 represents the relationship between retrieval time based on the number of users for IPFS-based Blockchain storage and Blockchain Storage. The retrieval time for the file will be based on the current users of the system. As the number of users increases, the time taken for storing the file will also increase. For IPFS-based Blockchain storage, the retrieval time even if there is more number of users, will be less.

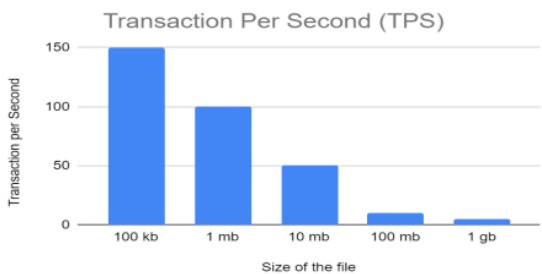

**Fig 12:- Transaction per second vs File Size**

      Figure 12 shows transaction per second vs File size. Transaction per second (TPS) is a metric used to measure the number of transactions processed by a system per second. The file size has an impact on the number of transactions per second. The number of transactions processed in a second is proportional to the size of the file.

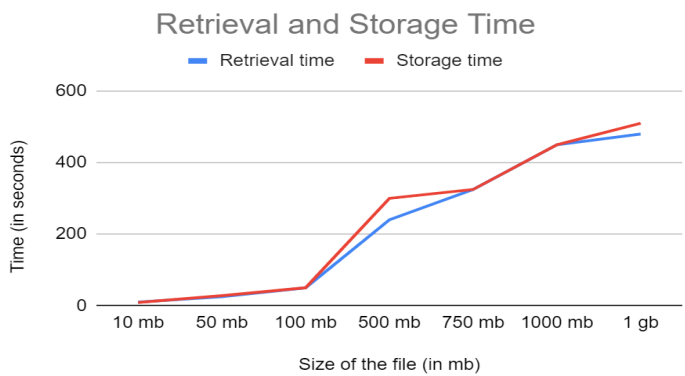

**Fig 13 :- Retrieval and Storage Time based on the size of the file.**

      Figure 13 shows the relationship between the size of the file, storage, and retrieval

time. The storage and retrieval time for the file will be based on the size of the file. As the file size increases, the time taken for storing and retrieving the file will also increase. For large files, the storage time will be quite large.

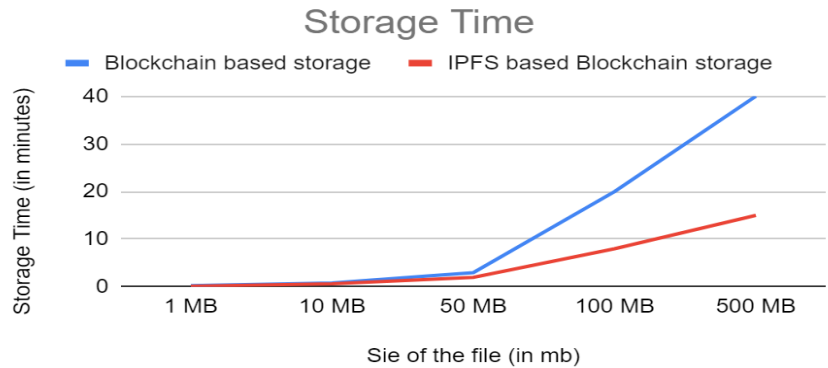

**Fig 14 :- Storage time for both Blockchain and IPFS based Blockchain based on file size.**

Figure 14 shows the comparison of storage time needed for Blockchain Storage and IPFS Based Blockchain storage based on the size of the file. The storage time needed for IPFS-based blockchain is less than the Blockchain-based storage. Therefore IPFS based blockchain storage takes minimum time to store a file.

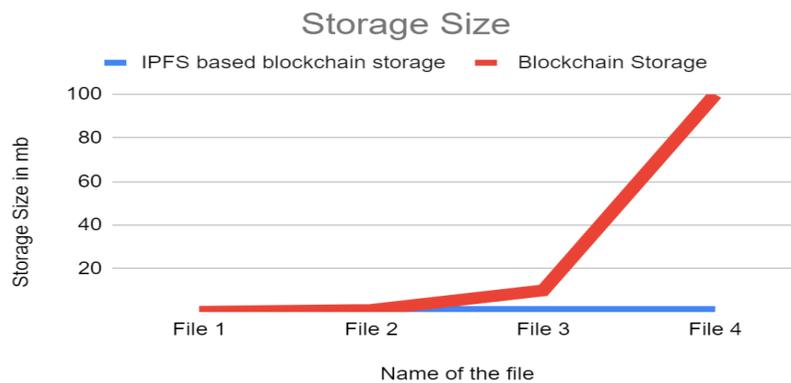

**Fig 15:- Storage size for Blockchain and IPFS-based Blockchain**

Figure 15 shows the comparison of storage size needed for Blockchain and IPFS-based Blockchain. The storage size of IPFS based Blockchain is small when compared to Blockchain storage.

## 5. Conclusion

The proposed method represents a significant advancement in enhancing the security and reliability of financial transactions in the digital age. By integrating smart contracts and blockchain technology, the system provides a transparent and tamper-resistant environment for conducting financial transactions, thereby reducing the risk of fraud and unauthorized access.Furthermore, the incorporation of FaceNet512 for improved face recognition and GMM for speech authentication adds an additional layer of security through multi-factor biometric authentication. This combination of cutting-edge technologies offers a robust defense against identity theft and illegal access, setting a new standard for secure financial transactions.Overall, the system's innovative approach leverages the strengths of various technologies to create a comprehensive solution that addresses the evolving challenges of security in the digital financial landscape. It paves the way for safer and more reliable financial interactions, ultimately benefiting individuals and businesses alike.

**Conflicts of interest**

All authors declare no conflict of interest in this paper.